\documentstyle[epsf,psfig,12pt]{article}
\newcommand{\be}{\begin{equation}}
\newcommand{\ee}{\end{equation}}
\newcommand{\ba}{\begin{eqnarray}}
\newcommand{\ea}{\end{eqnarray}}
\newcommand{\eqref}[1]{(\ref{#1})}

\newcommand{\1}{q_1}
\newcommand{\2}{q_2}
\newcommand{\3}{q_3}
\newcommand{\4}{q_4}
\newcounter{saveeqn}

\begin{document}

\centerline{\bf Time evolution of correlation functions and thermalization}
\vskip1cm
\renewcommand{\thefootnote}{\fnsymbol{footnote}}
\centerline{Gian Franco Bonini\footnote[1]{E-mail: bonini@thphys.uni-heidelberg.de} and Christof 
Wetterich\footnote[7]{E-mail: C.Wetterich@thphys.uni-heidelberg.de}}
\vskip0.5cm
\renewcommand{\thefootnote}{\arabic{footnote}}
\setcounter{footnote}{0}
\centerline{Institut f\"ur theoretische Physik, Universit\"at Heidelberg,\\}
\centerline{Philosophenweg 16, Heidelberg 69120, Germany\\}

\vskip 1cm
\centerline{July 27, 1999}

\setcounter{page}{0}
\thispagestyle{empty}

\vskip 1.5cm

\centerline{\bf Abstract}
\vskip0.2in
We investigate the time evolution of a classical ensemble of isolated periodic chains
of $O(N)$-symmetric anharmonic oscillators. Our method is based on   an exact evolution equation
for the time dependence of correlation functions. We discuss its solutions in
an approximation which  retains all contributions in next-to-leading
 order in a $1/N$ expansion and preserves time reflection symmetry. We observe effective 
irreversibility and approximate thermalization. At large time the 
system approaches stationary solutions in the vicinity of, but not identical to, thermal 
equilibrium. The ensemble therefore retains some memory of the initial condition beyond 
the conserved total energy. Such a behavior with incomplete thermalization is
referred to as "mesoscopic dynamics". It is 
expected for systems in a small volume. Surprisingly, we find that the nonthermal 
asymptotic stationary solutions do not change for large volume. This raises questions on 
Boltzmann's conjecture that macroscopic isolated systems thermalize.

\vskip1cm
\noindent HD-THEP-99-23  \\

\vfill

\eject

\baselineskip 24pt plus 2pt minus 2pt

\section{Introduction}
A central piece in our understanding of the dynamics of large statistical systems 
is Boltzmann's conjecture that an ensemble of isolated interacting systems approaches 
thermal equilibrium at large times. The asymptotic values of the correlation 
functions can then be computed from the resulting microcanonical equilibrium ensemble.
According to Boltzmann's conjecture their values only depend on the energy density of the
system or, equivalently, the temperature (with standard modifications in the presence of other
conserved extensive quantities). Besides the energy, all memory about the initial 
conditions is lost asymptotically - and also in practice if typical relaxation time scales
are not too large. The thermalization conjecture only applies to spatially extended 
systems in the limit of infinite volume. So far no proof for this hypothesis has been 
given. Correspondingly, the question of how effective irreversibility arises from 
microscopic equations which are invariant under time reflection, or from the time 
reversible Liouville equation, has not found a complete answer todate. This effective 
irreversibility is the basis of widely used effective equations (like the Boltzmann 
equation).

We want to address two issues by a direct study of the time dependence of the correlation
functions. The first concerns isolated systems with a finite number of degrees of freedom,
 corresponding to a finite volume $V$. No equilibration is expected for microscopic 
systems of only a few degrees of freedom. Recently, this has been demonstrated explicitly 
for the correlation functions of coupled anharmonic oscillators \cite{lb-cw1} \footnote{The issue
would be completely different, of course, if the system were coupled to an external heat 
bath.}. Let us assume for a moment  that  thermalization occurs for macroscopic systems. 
Then a smooth transition between 
the two extremes requires that there must be some intermediate 
volume or number of degrees of freedom where thermalization is incomplete, {\em i.e.} 
the ensemble retains some memory of the initial conditions. We may call the associated 
time evolution of the correlation functions ``mesoscopic dynamics''. Mesoscopic dynamics 
is characterized by partial effective irreversibility on one side, {\em i.e.} many 
details of the initial conditions get lost for asymptotic times. On the other hand, this 
``loss of memory'' is not complete (as for strictly thermalizing systems) so that partial 
information about the initial state can be recovered even after an arbitrarily long time. 

The second question concerns the validity of Boltzmann's conjecture for systems with a 
large volume. It is conceivable that thermalization remains incomplete even for macroscopic isolated
systems. In this case some characteristics of mesoscopic dynamics would
survive in the infinite   volume limit. We emphasize that
mesoscopic dynamics is always relevant in an appropriate volume range.
Our second question therefore asks if and how the asymptotic loss of memory becomes complete as the
volume becomes macroscopic. Within our approximations we find that certain
features of mesoscopic dynamics remain 
present for isolated systems in the large volume limit!

Our investigation is based on an exact evolution equation for the time dependent 
effective action \cite{formalism} which is the generating functional of the equal time one 
particle irreducible (1PI) correlation functions. The direct study of the time evolution of 
the correlation functions circumvents the calculation of the time dependence of the 
probability distribution or density matrix. For classical systems the exact evolution 
equation is equivalent to the BBGKY hierarchy \cite{BBGKY}. A study of the 1PI
correlation functions offers, however,  new possibilities of systematic truncations. In 
particular, thermal equilibrium is now present as a stationary solution at every step of the
 truncation. Also the generalization to quantum statistics is straightforward and surprisingly
 simple \cite{cw:hn}. An investigation of the general structure of the exact evolution equation for the time dependent effective action reveals many new stationary solutions besides thermal equilibrium \cite{lb-cw2}. 
The question of their dynamical role is part of the scope of
 this paper.

We will concentrate here on a particular example, namely a periodic chain of coupled 
anharmonic oscillators with $O(N)$ symmetry. The Hamiltonian

\begin{equation}
H=\int_0^l{dx\left[\frac{1}{2}p_a(x)p_a(x)+\frac{1}{2}q_a(x)(m^2-\Delta) q_a(x)
+\frac{\lambda}{8N}\left(q_a(x)q_a(x)\right)^2\right]}
\end{equation}
describes an interacting time reversible system with microscopic time evolution given by
\ba
\partial_t q_a(x)&=&p_a(x)\nonumber\\
\partial_t p_a(x)&=&-(m^2-\Delta)q_a(x)-\frac{\lambda}{2N}q_b(x)q_b(x)q_a(x).\label{eq:micro}
\ea

We may assume that the oscillators sit 
on discrete lattice sites with distance $a$ such  that high momenta are cut off. Then 
$\Delta$ is an appropriate discretized Laplace operator involving neighboring sites.
The index $a=1...N$ counts the oscillators at a given site and $H$ preserves 
$O(N)$ symmetry. Practical examples for small $N$ may be found in the form of ring-sized 
molecules with discrete translational symmetry along the ring. For example, the $q_a$ may describe 
 displacements from the equilibrium position (with $p_a$ the associated momenta). The 
limit of large $l$ also describes large linear molecules if boundary effects from the 
ends can be neglected. Another interesting limit is $\lambda \rightarrow \infty$ for 
fixed $m^2/\lambda =-1$. This imposes the nonlinear constraint $q_aq_a=1$ and describes 
classical bosonic spin chains ($N=3$ for spin one). The length $l$ of the chain plays the role of the volume 
in three dimensional systems. For $l=0$ we are left with an ensemble of simple 
$N$-component anharmonic oscillators. This case has been studied in detail in  
\cite{lb-cw1}. No thermalization is possible since infinitely many conserved correlation 
functions keep the memory of the initial condition and obstruct thermalization. The 
conserved cumulants are simply related to powers of the conserved energy and squared 
$O(N)$-angular momentum $L^2$, {\em i.e.} $<E^r(L^2)^s>$ does not change in time for 
arbitrary $r$ and $s$.
If Boltzmann's conjecture is true, asymptotic thermalization governs the behavior for 
$l\rightarrow \infty$. In this case we conclude that there must be a range of $l$ with 
mesoscopic dynamics, describing the transition between the limits $l\rightarrow 0$ and 
$l\rightarrow \infty$. On the other hand, if Boltzmann's conjecture does not hold for 
isolated systems of this type some features of mesoscopic dynamics are  expected to be relevant also 
for $l\rightarrow \infty$.

Our system is also a prototype for classical and quantum field theories. From this point
 of view it describes a one-dimensional $O(N)$-symmetric scalar field theory. There is 
no conceptual problem in its generalization to three dimensions where it would be 
relevant for cosmology ({\em e.g.} inflation and parametric resonance or dynamical scalar 
fields playing a role in late cosmology), for particle physics ({\em e.g.} pions in 
heavy ion collisions) or for many systems in statistical mechanics.

Although most practical applications are for small $N$, we also discuss the limit of 
large $N$. The reason is that one of our truncation schemes is a systematic expansion 
in powers of $1/N$. The leading order in the $1/N$ expansion has been discussed by 
various groups with different methods \cite{1overN}, \cite{lb-cw2}. 
In this first approximation 
infinitely many conserved quantities preclude thermalization \cite{lb-cw2}. We include 
here all contributions in next to leading order in $1/N$. In particular, this  includes 
scattering in three dimensional field theories. Particle numbers for individual momentum modes are 
no longer conserved and 
there is no immediately visible obstruction to thermalization anymore.

We concentrate in this paper on classical statistics. This has the advantage that our 
results can easily be compared with other methods. In particular, it should be feasible to
 solve the microscopic equations numerically with 
given initial values and then take averages over an ensemble of initial values. In this 
way the equal time correlation functions discussed in this paper can be directly measured
 at any later time. The generalization to quantum statistics is straightforward in our 
approach and will be postponed to a subsequent paper.

We investigate homogeneous (translation invariant) and $O(N)$ symmetric ensembles. 
Individual members of these ensembles do not, of course, share this high degree of symmetry. 
For generic initial conditions the solution of the microscopic evolution
equation (\ref{eq:micro}) is highly 
inhomogeneous and shows no $O(N)$ symmetry. The high symmetry of the ensemble
only means that we  weight the initial conditions according to  
a probability distribution that exhibits this symmetry. In practice, there 
is actually no
need to specify the probability distribution at the initial time $t_0$
explicitely.  It is often more 
effective to specify the correlation functions at $t_0$. These will constitute
the initial data for  our 
differential flow equations.   In the present paper we mainly consider
gaussian initial perturbations from equilibrium, where all 1PI 
$n$-point functions except the two-point functions are equal to their thermal
values. 

\begin{figure}
\vspace{-8cm}
\centerline{\psfig{file=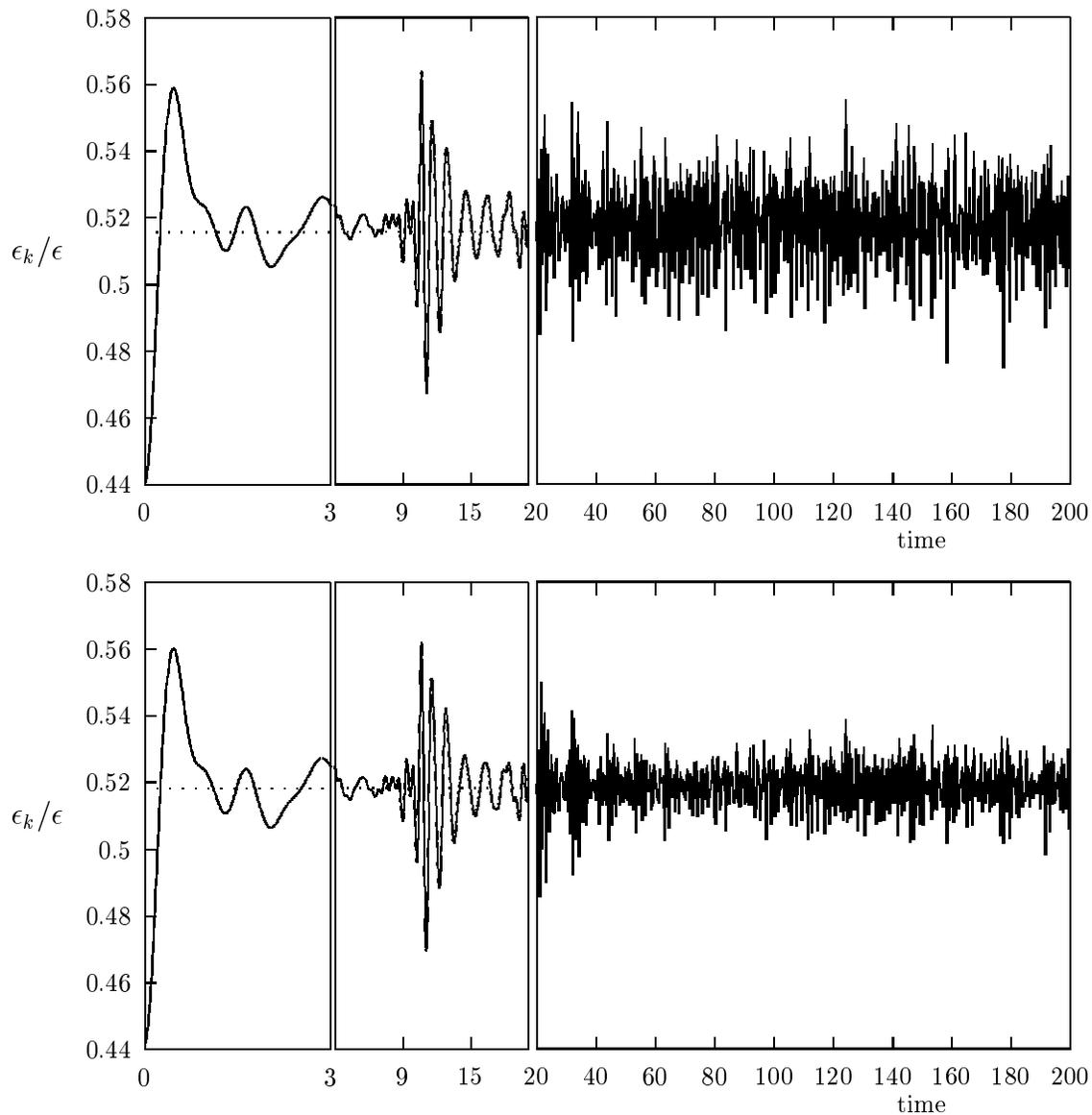}}
\vspace{-9cm}

\caption{Energy ``equipartition''. We show $\epsilon_k/\epsilon$ as a function of time, in the leading (upper
  panel)  and next-to-leading $1/N$ approximations, for $\lambda=2$, $N=20$,
 $\Lambda=5$, and $l=20.1$. The system  is initially  displaced from
 equilibrium  according to a gaussian  perturbation in $B(q)$ ($\beta=0.5$, 
$D_B=0.25$,
 $\Delta_B=0.469$, $q_B=2.5$; cf. Eq.(\protect\ref{eq:perturbation})). 
Horizontal lines correspond to  thermal equilibrium for $T=2$.}
\label{Evst_intro}
\end{figure}

\begin{figure}
\vspace{-12cm}
\centerline{\psfig{file=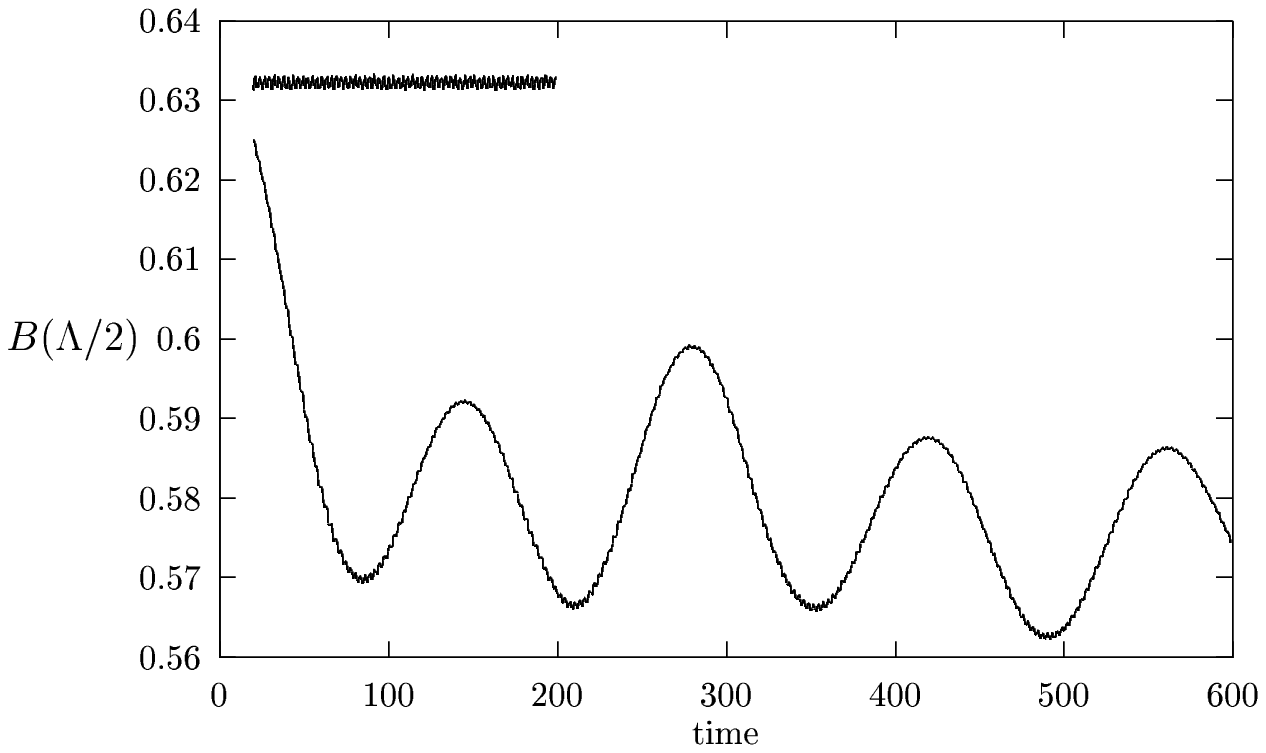}}
\vspace{-10cm}
\caption{Time evolution of the two-point correlation function. We plot $B(\Lambda/2)$ for the same system as in 
Fig.\protect\ref{Evst_intro}, in the leading (upper curve) and next-to-leading $1/N$
approximations. 
The plotted values are averages over the time interval $[t-20,t]$. Note that 
the equilibrium value for the corresponding energy is $B(\Lambda/2)=0.5$. In the 
leading $1/N$ approximation $B(q)$ stays esssentially constant for arbitrarily long time.} 
\label{Bvst_intro}
\end{figure}

We find effective irreversibility as a property of the solutions of our time
reversible flow  equation. In a 
wider sense this is due to the existence of attractive fixed point
solutions. In our context an  attractive fixed 
point does not necessarily mean that all solutions for a given class of
initial ensembles  asymptotically reach 
this fixed point. The characteristic behavior for large $t$ is rather
characterized  by high-frequency oscillations of the correlation
 functions around time averaged stationary mean values. The approach to the
 stationary behavior for the time averaged 
correlation functions shows three characteristic features:
\begin{itemize}
\item{First we find a fast initial irreversible behavior on typical
    microscopic  time scales between the inverse 
momentum cutoff $\Lambda ^{-1}=a/\pi$ (with $a$ the lattice distance) and the inverse mass
$m^{-1}$. An example is given in Fig.\ref{Evst_intro}  where we plot
the time evolution of the ratio between kinetic energy
and total energy
for an out-of-equilibrium system. Comparison with the leading order in the
$1/N$ expansion reveals no qualitative
difference. We conclude that this first period of irreversibility is not
related to scattering but rather
described by dephasing \cite{1overN}, \cite{lb-cw2}. A ``rough thermalization'' takes already
place at this very early stage.}
\item{
The first stage of rapid ``rough thermalization'' does not bring the two-point
functions near the equilibrium values.
In Fig.\ref{Bvst_intro} we display the evolution of the time-averaged two-point function 
$B(q)$
which characterizes the gaussian part of the probability distribution for the
momenta 
$p_a(q)=\int dx\; e^{-iqx}p_a(x)$  by $<p_a(q)p_b(q')>\;=2\pi\delta(q+q')\delta_{ab}G^{\pi\pi}(q)$,
$G^{\pi\pi}(q)=B^{-1}(q)(1-C^2(q)/(A(q)B(q)))^{-1}$ (see below). 
In thermal 
equilibrium one expects the Maxwell velocity distribution with
$B(q)=\beta=1/T$  independent of $q$ and 
$C(q)=0$. One observes that an initially disturbed $B(q)$ approaches
a stationary value only on time  scales much larger than
$m^{-1}$. ``Scattering'' is essential for this aspect of irreversibility. This
can be seen by a comparison with 
the leading $1/N$ behavior where $B(q)$ oscillates around a time independent
value  for every $q$. In leading
order $1/N$ no energy is exchanged between the different Fourier modes. This
explains why time averaged values 
for $B(q)$ are stationary from the beginning and therefore cannot equilibrate.}
\item{The exchange of energy between different Fourier modes in
    next-to-leading  order in the $1/N$ expansion 
drives the time averaged velocity distribution toward a
stationary  value. It may be a surprise
that in general this stationary value differs from thermal equilibrium
with uniform $B(q)=\beta$.  The implications
of this finding will be discussed in the last section.}

\end{itemize}

\section{The method}
Our investigation is based on the time-dependent effective action
\cite{formalism}, which  generates the equal-time 1PI correlation functions.
We consider  an $N$-component $(1+1)$-dimensional scalar
$\phi^4$ theory and ensembles which are invariant under internal $O(N)$ transformations,
spatial translations, and reflection. Our truncation retains all 1PI $n$-point
 functions up to $n=4$ and omits 1PI vertices with $n\geq 6$. One should notice that this 
still includes connected $n$-point functions with arbitrary $n$. In this approximation 
the effective action of our model becomes:

\begin{eqnarray}
\Gamma [\phi,\pi;t]&=& {1 \over 2} \int {d^D q \over (2 \pi)^D } 
\Bigl\{  A(q)\phi_a^*(q)\phi_a(q)  +   B(q) \pi_a^*(q) \pi_a(q) 
\nonumber \Bigr. \\ && \Bigl. +  2 C(q) \pi_a^*(q)
\phi_a (q) \Bigr\} 
+ {1\over 8N} \int {d^D q_1 \over (2 \pi)^D} {d^D q_2 \over (2 \pi)^D} 
{d^D q_3 \over (2 \pi)^D}  \nonumber \\ 
&& \Bigl\{ 
u(q_1,q_2,q_3)  \phi_a(q_1) \phi_a(q_2) \phi_b(q_3) 
\phi_b(-q_1-q_2-q_3) \Bigr. \nonumber \\
&& \Bigl. 
+ v(q_1,q_2,q_3) \pi_a(q_1) \phi_a(q_2) \phi_b(q_3) 
\phi_b(-q_1-q_2-q_3) \Bigr. \nonumber \\
&& \Bigl. 
+w(q_1,q_2,q_3)\pi_a(q_1) \pi_a(q_2) \phi_b(q_3) 
\phi_b(-q_1-q_2-q_3) \Bigr. \nonumber \\
&& \Bigl. + s(q_1,q_2,q_3) \left[\pi_a(q_1) \pi_b(q_2) \phi_a(q_3) 
\phi_b(-q_1-q_2-q_3) \right. \Bigr. \nonumber \\
&& \Bigl. \left. \qquad - \pi_a(q_1) \pi_a(q_2) \phi_b(q_3) 
\phi_b(-q_1-q_2-q_3) \right]  \label{e4} \Bigr.  \\
&& \Bigl. + y(q_1,q_2,q_3) \pi_a(q_1) \pi_a(q_2) \pi_b(q_3) 
\phi_b(-q_1-q_2-q_3) \Bigr. \nonumber \\
&& \Bigl. +  z(q_1,q_2,q_3) 
\pi_a(q_1) \pi_a(q_2) \pi_b(q_3)  \pi_b(-q_1-q_2-q_3) \Bigr\}, 
\nonumber 
\end{eqnarray}
where the ``couplings'' $A(q)$, $u(q_1,q_2,q_3)$, etc. depend on time.
The 1PI $n$-point functions are obtained by taking derivatives of $\Gamma$
with respect to $\phi$ and $\pi$, the second derivative being the inverse
propagator.  For example, the connected two-point function for $q_a$ reads
\begin{equation}
<q_a(x)q_b(y)>_c=G(x-y)\delta_{ab}=\int \frac
{d^Dq}{(2\pi)^D}G(q)e^{iq(x-y)}\delta_{ab},
\end{equation}
where 
\begin{equation}
G(q)=\frac{B(q)}{A(q)B(q)-C^2(q)}.
\label{eq:G}
\end{equation}

The time evolution of $\Gamma$ induced by Eq.(\ref{eq:micro}) is dictated by the nonlinear evolution operator

\begin{eqnarray}
\partial_t \Gamma[\phi, \pi;t] = - \left({\cal L_{ \rm cl}} 
+ {\cal L_{\rm q}} \right) \Gamma[\phi, \pi;t],
\label{e1}
\end{eqnarray}

where ($\psi_i\equiv (\phi_a,\pi_a)$)

\begin{eqnarray}
{\cal L}_{\rm cl} &=& \int d^D x \left\{ \pi_a(x) 
{\delta \over \delta \phi_a(x)} + \phi_a(x)\left(\nabla^2 - m^2 \right)
{\delta \over \delta \pi_a(x)} \nonumber \right. \\ 
&&- \left. {\lambda \over 2N} 
\left[\phi_b(x) \phi_b(x) \phi_a(x) +  \phi_a(x) 
G^{\phi \phi}_{bb}(x,x) +   2 \phi_b(x) G^{\phi \phi}_{ba}(x,x) 
\nonumber  \right. \right. \\ 
&& -  \left. \left. 
\int d^Dx_1  d^Dx_2  d^Dx_3 G^{\phi \psi}_{ai}(x,x_1)
G^{\phi \psi}_{bj}(x,x_2) \nonumber \right. \right. \\
&& \times \left. \left.
G^{\phi \psi}_{bk}(x,x_3) {\delta^3 \Gamma \over 
\delta \psi_i(x_1) \delta \psi_j(x_2) \delta \psi_k(x_3)} \right ] 
{\delta \over \delta \pi_a(x) } \right\} , 
\ea
and
\ba
{\cal L}_{\rm q} 
&=& {\lambda \over 8N} \hbar^2\int d^D x~ \phi_a(x)
{\delta \Gamma \over \delta \pi_b(x)} 
{\delta \Gamma \over \delta \pi_b(x)} 
{\delta  \over \delta \pi_a(x)},
\label{e3}
\end{eqnarray}
with
\be
\left[G^{-1}\right]^{\psi\psi'}_{ab}(x,y)=
\frac{\delta^2\Gamma}{\delta\psi_a(x)\psi'_b(y)},
\ee
and $G$ of (\ref{eq:G}) corresponding to $G^{\phi\phi}$. 

The exact flow equations for the two-point functions follow from taking the second derivatives
of Eq.(\ref{e1}) with respect to $\phi$ and $\pi$ at $\phi=\pi=0$:

\ba
\dot A(q)&=&2\omega^2(q)C(q)\nonumber\\
\dot B(q)&=&-2C(q)-\frac{2}{N}\gamma(q)B(q)\label{fleqfirst}\\
\dot C(q)&=&-A(q)+\omega^2(q)B(q) -\frac{\gamma(q)}{N} C(q),\nonumber
\ea
where

\ba
\omega^2(q)&\equiv&q^2+m^2+\lambda\frac{(N+2)}{2N}\int_pG(p)\nonumber\\
&&-\frac{\lambda(N+2)}{8N^2}\int_{q_1,q_2}G(q_1)
G(q_2)G(q-q_2-q_1)\cdot \nonumber\\
&&[4u(q_1,-q,q_2)-c(q_1)c(q_2)c(q-q_1-q_2)y(q-q_1-q_2,q_1,q_2)
\nonumber \\
&&-c(q_1)\left[2v(-q_1,-q_2,q)+v(-q_1,q,-q_2)\right]\nonumber \\
&&+2c(q_2)c(-q_1-q_2+q)w(q-q_1-q_2, q_2,q_1)]\nonumber 
\ea

\ba
\gamma(q)&\equiv&\frac{\lambda(N+2)}{8N}\int_{q_1,q_2}G(q_1)
G(q_2)G(q-q_2-q_1)\cdot \nonumber\\
&&[v(q,-q_1,-q_2)-4c(q_1)c(q_2)c(q-q_1-q_2)z(-q_1,q,-q_2)\nonumber \\
&&-2c(q_1)w(-q_1,q,-q_2)\nonumber\\
&&+c(q_2)c(q-q_1-q_2)[y(q-q_1-q_2,q_2,-q)+2y(-q,q_1,q-q_1-q_2)]]\nonumber
\ea

\ba
c(q)&\equiv& \frac{C(q)}{B(q)}.
\ea

Similarly, the flow equation for the quartic coupling $u$ reads
\footnote{We display here only one of the six flow equations for the 4-point 
couplings.
The remaining five equations can be found in Appendix A.}:

\ba
   \dot u(\1,\2,\3)&=&\left[\omega^2(\1)v(\1,\2,\3)+4\lambda C(\1)
        -4\lambda C(\2)(S_1(\1+\2,\3)\right.\nonumber\\ &&\left.+ S_2(\2+\3,\1))
        -\lambda \hbar^2C(\1)C(\2)C(\3)\right]_{SYM},
\label{fleqlast}
\ea
where the subscript $SYM$ implies symmetrization with respect to the
appropriate permutations of $\1$, $\2$, $\3$ and $\4=-(\1+\2+\3)$. Here
we have introduced the momentum integrals
\ba
S_1(\1,\2)&\equiv&\frac{1}{2N}\int_q G(q)G(q+\1)\cdot\nonumber\\
&&\left[(N+2)u(q+\1,-q,\2)+2u(\2,-q,q+\1)
\right.\nonumber\\
&&-\frac{1}{2}
c(q)((N+2)v(-q,q+\1,\2)+2v(-q,\2,q+\1))\nonumber\\
&&+\frac{1}{2}
c(q+\1)c(q)((N+2)(w(-q,q+\1,\2)-\frac{1}{2}s(-q,q+\1,\2)\nonumber\\
&&\left.-\frac{1}{2}s(q+\1,-q,\2))
   +s(-q,q+\1,\2))\right]\nonumber\\
S_2(\1,\2)&\equiv&\frac{1}{2N}\int_q G(q)G(-q-\1)
\left[  4u(-q,\2,q+\1)\right.\nonumber\\
&&+c(q)c(-q-\1)s(q+\1,-q,-\1-\2)\nonumber\\
  &&\left.-c(q)(v(-q,\2,q+\1)+v(-q,-\1-\2,q+\1))\right].
\ea

The flow equations for the quartic couplings $u$, $v$, etc. are not exact
since we have truncated the contributions from 1PI 6-point functions. We
furthermore have omitted the two-loop contribution to
the evolution of the quartic couplings. Our approximation may be viewed as the
 second order in a weighted loop 
expansion where the evolution of every 1PI $2m$-point function is computed in
$(n_L+1-m)$-loop order ({\em i.e.} two loops for the two-point function, one
loop  for the four-point function).
It is easy to convince oneself that this expansion retains systematically all
contribution in order $N^{1-n_L}$. In our case  it also includes (incompletely)
terms of order $1/N^2$. For comparison we employ  a 
second systematic expansion, namely the $1/N$ expansion, where all terms of
order $1/N^2$ are omitted \footnote{With the exception of the subleading  contributions to
  the evolution of the 4-point couplings that are contained in 
  $\omega^2$ and $\gamma$. These have to be retained in order to ensure exact
  energy conservation.} in the flow equations. 

The flow equations conserve exactly the energy density $\epsilon=E/l$:
\ba
\epsilon&=&\frac{N}{2}\int_q \left\{ B^{-1}(q)+G(q)\left[q^2+m^2+c^2(q)+\frac{N+2}{4N}\lambda
  \int_p G(p)\right]\right\} \label{eq:energy}\\
&&-\frac{N+2}{8N}\lambda\int_{\1,\2,\3} G(\1)G(\2)G(\3)G(\4)
    \left[ u(\1,\2,\3)- v(\1,\2,\3)c(\1) \right.\nonumber\\
        &&\left.+w(\1,\2,\3)c(\1)c(\2)-y(\1,\2,\3)c(\1)c(\2)c(\3)\right.\nonumber\\
        &&\left.+ z(\1,\2,\3)c(\1)c(\2)c(\3)c(\4) \right],\nonumber
\ea
whereas the squared $O(N)$ "angular momentum" density
\ba
&&\frac{L^2}{l}=N(N-1)\int_{\1}G(\1)B^{-1}(\1)\cdot\label{eq:momentum}\\
&&\left\{1-\frac{1}{4N}\int_{\2} G(\2)B^{-1}(\2)(2w(\1,\2,-\2)-2s(\1,\2,-\2)-
   s(\1,\2,-\1))\right\}\nonumber
 \ea
is conserved only up to relative corrections of order $1/N^2$.
Additional independent conserved quantities of the form $<E^r(L^2)^s>-<E>^r<(L^2)>^s$ are 
suppressed by inverse powers of $N$. They are  not conserved by the truncated 
equations. The kinetic energy density $\epsilon_k=\frac{N}{2}\int_q \left\{ B^{-1}(q)+G(q)c^2(q)\right\}$ 
is, of course, not separately conserved.

We have solved the classical flow equations ($\hbar =0$) numerically for a discretized 
system with   $N_l$ points  and an ultraviolet cutoff $\Lambda =$
a few times 
$m$, using a standard fourth-order Runge-Kutta algorithm which has the
property of being exactly time-reversible. We only consider here positive
$m^2$ and set the mass scale by $m=1$. For a typical cutoff $\Lambda=5$ and 
$N_l=32$, the length of the chain is $l=N_la=\pi N_l/\Lambda\simeq20.1$. 
\section{Equilibrium properties}
\begin{figure}
\vspace{-8cm}
\centerline{\psfig{file=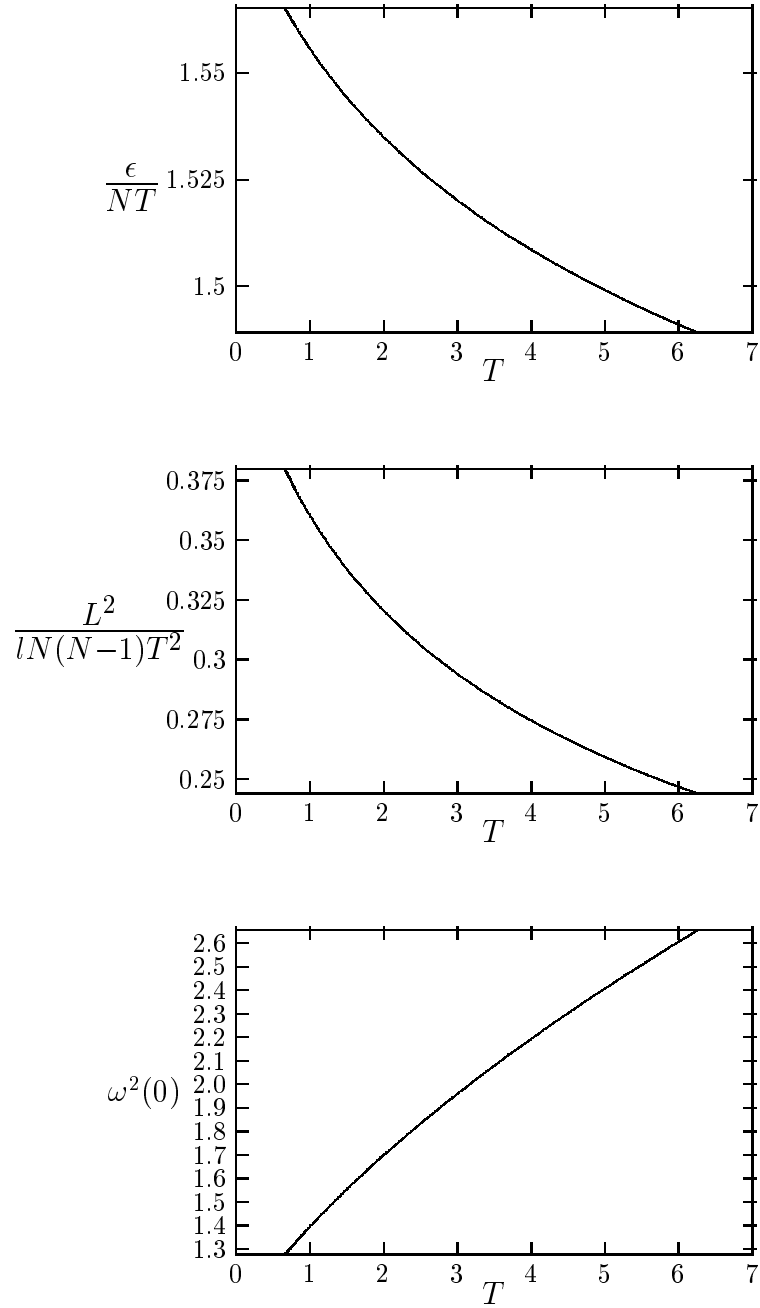}}
\vspace{-6.5cm}
\caption{Thermodynamic equilibrium properties. We show $\epsilon/NT$, $L^2/(lN(N-1)T^2)$, and $\omega^2(0)$ as functions of $T$, for a system
with $N=20$, $\lambda=2$, $\Lambda=5$, and $l=20.1$.}
\label{thermal1}
\end{figure}

As a first step in the  numerical analysis, we compute  the
classical 
thermal equilibrium 
configuration (defined by the conditions: $C=v=w=s=y=z=0$, $B=\beta$)
for 
different values of the parameters. In our approximation this 
corresponds to the solution of the Schwinger-Dyson equations
for $A$ and $u$ that follow from the requirement $\partial_t \Gamma=0$.
For $\lambda/N\ll1$ it is possible to 
derive the thermal values of $A$ and $u$ 
iteratively as  power series in $\lambda/N$. In general, however, 
this method fails, and we found it simpler to use the flow equations themselves, 
starting from the $N\rightarrow \infty$ thermal fixed point, letting the
system  evolve 
for a while ($\Delta t > m^{-1}$), taking time averages of the correlation
functions, adjusting them in accordance
with the thermal fixed point constraints, and repeating the procedure until 
stationary behavior with the desired accuracy 
was obtained.
We were thus able to obtain configurations that were thermal to a very good 
approximation ($\Delta B/B <0.001$). 
In Fig. \ref{thermal1} we display the energy density and the squared angular
momentum 
as functions of the temperature. We also show the frequency $\omega_{eq}(0)$
which is related in 
equilibrium to a (partially) renormalized temperature-dependent mass by
$\omega^2_{eq}(q)=A_{eq}(q)T=T/G_{eq}(q)=m_R^2+Z(q)q^2$.

\section{Dephasing and scattering}

Next we discuss the evolution of ensembles that are initially not in  thermal 
equilibrium. We opted for  gaussian perturbations from the thermal state with
initial two-point functions:
\ba
A_0(q)&=&C_A\left(A_{eq}(q)+D_A\left[e^{-\frac{(q-q_A)^2}{\Delta_A^2}}+
e^{-\frac{(q+q_A)^2}{\Delta_A^2}}\right]\right)\nonumber\\
B_0(q)&=&C_B\left(\beta+D_B\left[e^{-\frac{(q-q_B)^2}{\Delta_B^2}}+
e^{-\frac{(q+q_B)^2}
{\Delta_B^2}}\right]\right)\label{eq:perturbation}\\
C_0(q)&=&D_C\left[e^{-\frac{(q-q_C)^2}{\Delta_C^2}}+e^{-\frac{(q+q_C)^2}
{\Delta_C^2}}\right].\nonumber
\ea
The constants $D_A$, $D_B$, $D_C$, $q_A$, $q_B$, $q_C$, $\Delta_A$,
$\Delta_B$, $\Delta_C$ are arbitrary, whereas  $C_A$ and $C_B$ are tuned so that the perturbed 
system has the same $E$ and $L^2$ as the  unperturbed thermal equilibrium ensemble.
We also use superpositions of gaussian perturbations with the property that
the initial deviations $D_A$, $D_B$  for small and large $q^2$ are small.
 
In order to assess the importance of ``scattering'' for the equilibration of different
physical quantities, we first compare the results obtained by using the full 
equations (\ref{fleqfirst})-(\ref{fleqlast}) with those obtained by 
keeping only the leading terms in $1/N$ ({\em i.e.} neglecting all 4-point functions).
We repeat here that in leading order $1/N$ 
scattering is absent and only kinetic dephasing can induce
a smoothening and averaging out of the perturbation.

\begin{figure}
\vspace{-6cm}
\centerline{\psfig{file=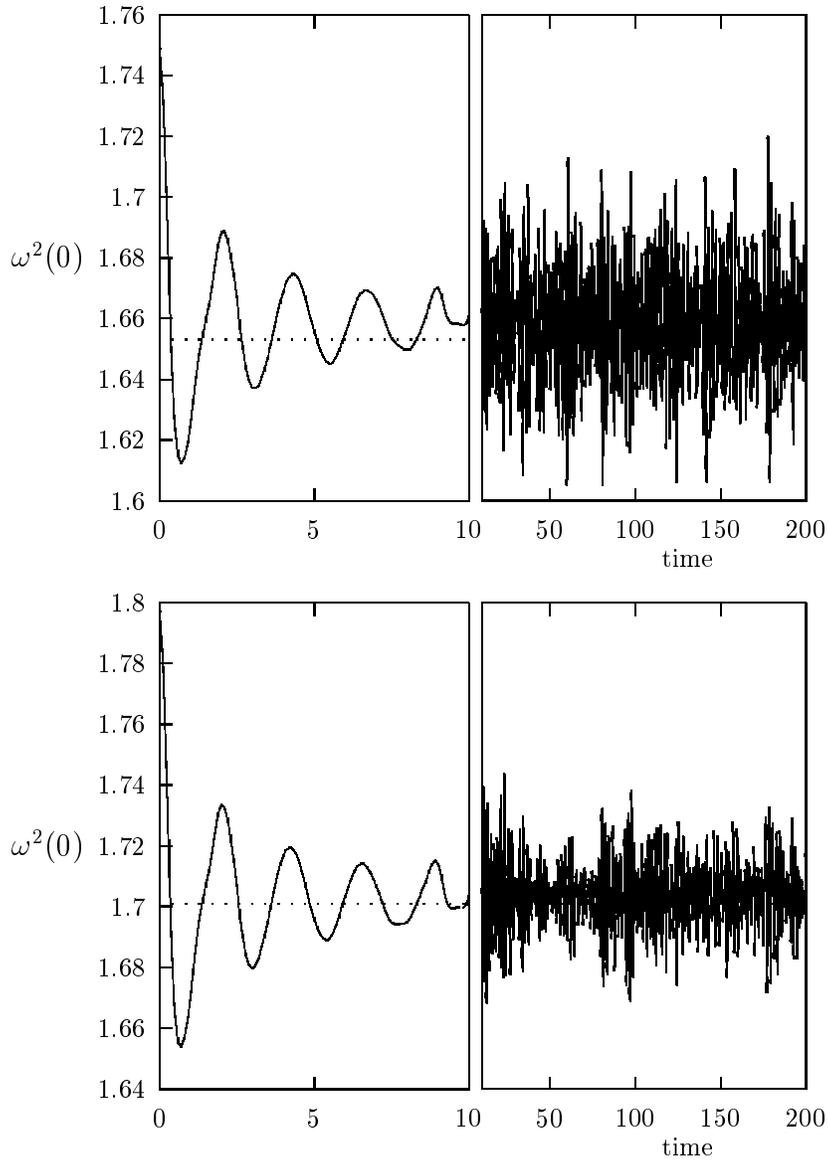}}
\vspace{-9.5cm}
\caption{Evolution of the frequency $\omega^2(0)$, in the leading (upper) and next-to-leading
  $1/N$ approximations  (same parameters as in  Figs.1-2). Comparison with the equilibrium value (also shown) indicates the ``more thermal'' behavior due to the inclusion of scattering, see also Fig.\ref{thermal1} for $T=2$.
}
\label{figure:5_1_2_20om}
\end{figure}

\begin{figure}
\vspace{-6cm}
\centerline{\psfig{file=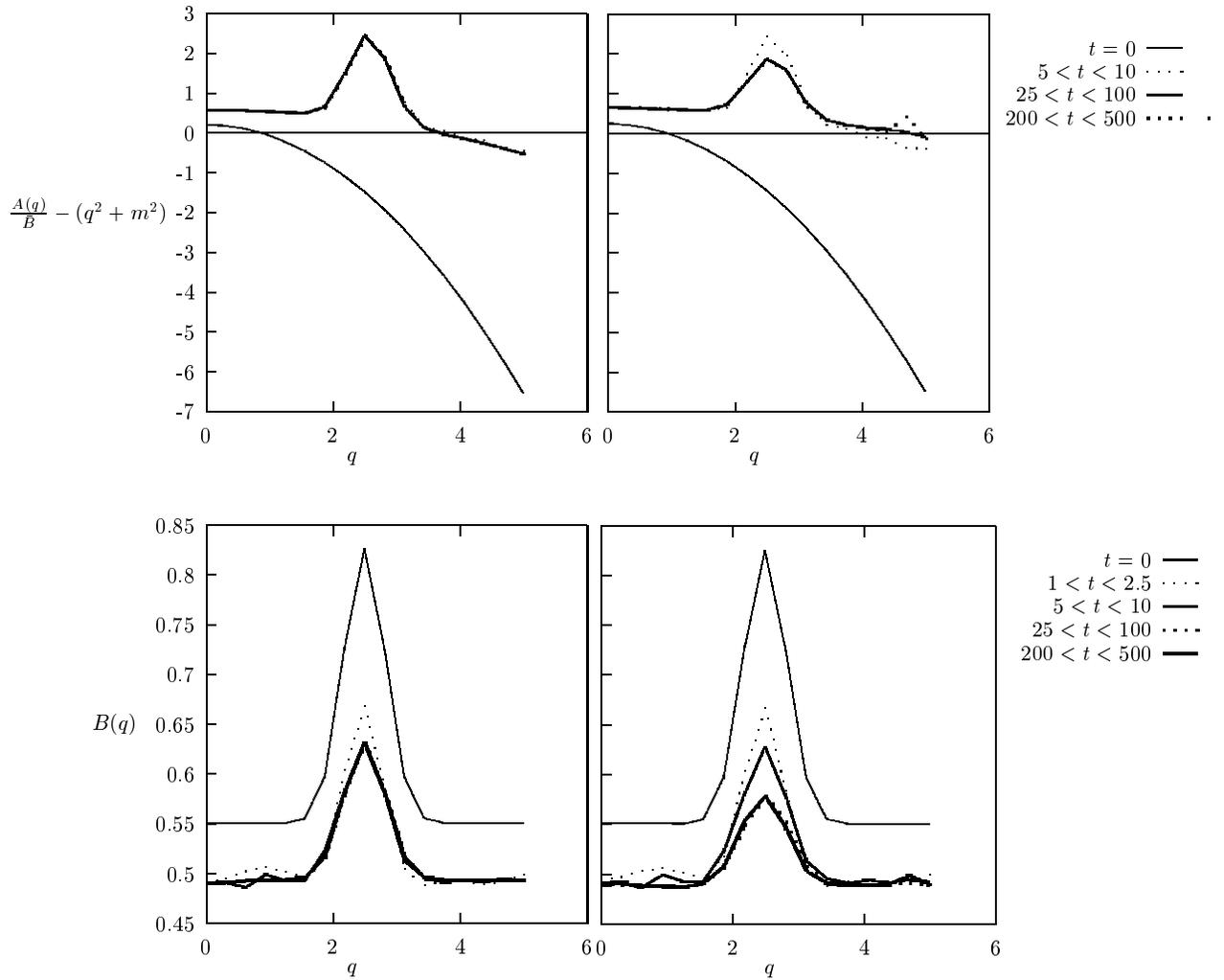,width=18cm}}
\vspace{-7cm}
\caption{A comparison between the leading-order (left) and next-to-leading-order 
(right) evolutions of the two-point correlation functions.
The first row shows $A(q)/(\sum_{q'}B(q')/N_l)-(q^2+m^2)$, which is 
a measure for the deviation of the inverse propagator from the classical value. The second row
gives  $B(q)$, or the deviation from the Maxwell velocity distribution $B(q)=\beta$. The correlation
functions are averaged over  various time intervals. The initial values at $t=0$ are also shown.}
\label{figure:5_1_2_20ab}
\end{figure}

\begin{figure}

\vspace{-8cm}
\centerline{\psfig{file=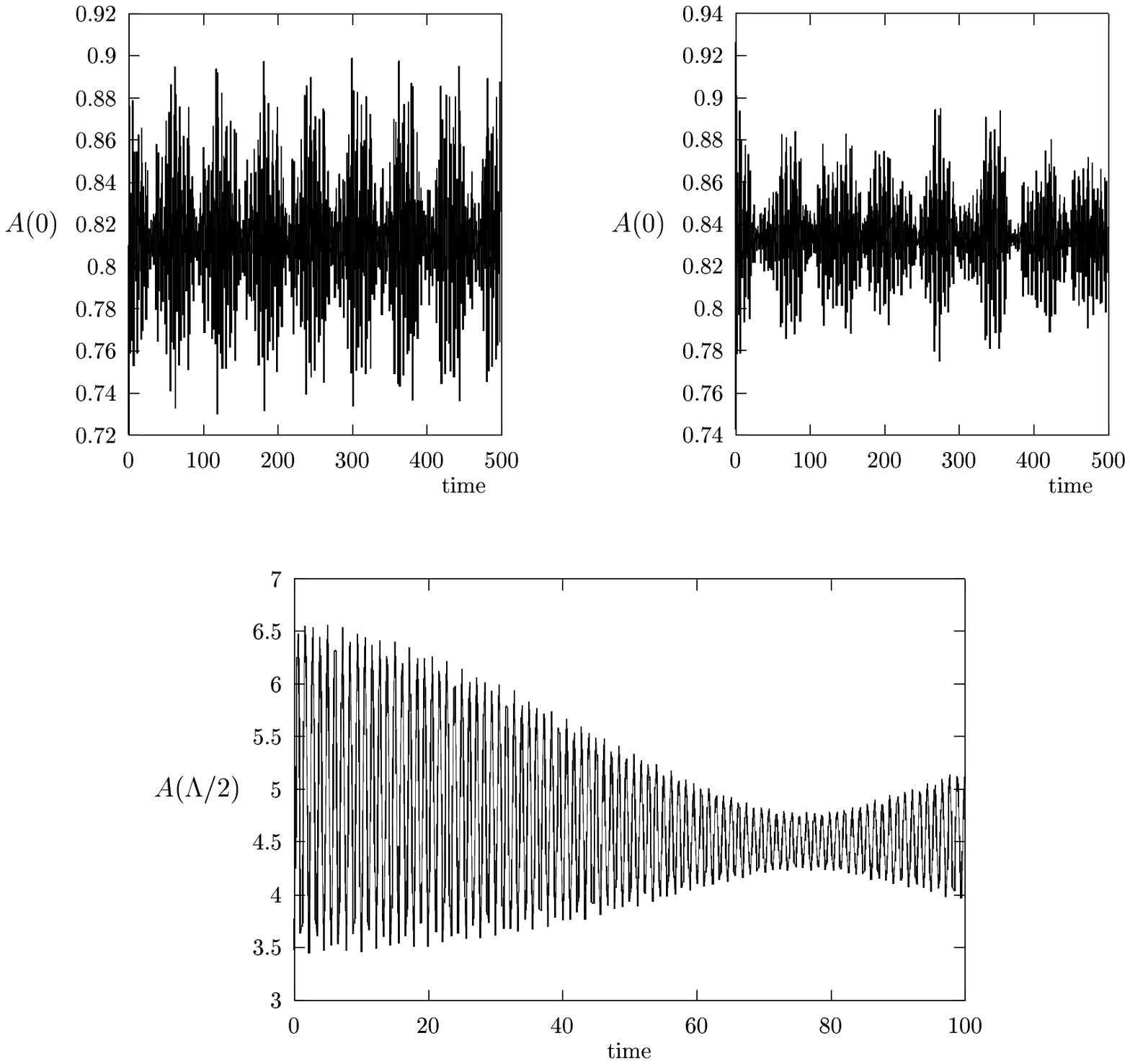}}
\vspace{-7.5cm}
\caption{Full time evolution of the two-point functions $A(0)$ (both leading and next-to-leading order) 
 and $A(\Lambda/2)$ (next-to-leading only), without averaging.}
\label{figure:5_1_2_20avst}
\end{figure}

As we can see in Fig.\ref{Evst_intro}, 
even in the absence of interactions, 
energy equipartition is  achieved to a very good approximation. Also $\omega(0)$ 
equilibrates approximately (Fig.\ref{figure:5_1_2_20om}). The individual correlation functions 
$A(q)$ and $B(q)$, however,  do not equilibrate in the absence of scattering (Figs.\ref{Bvst_intro},
\ref{figure:5_1_2_20ab}). 
They oscillate around constant values. 
When the effect of the time-dependent four-point functions is  added
 the picture changes.
We now see that the original perturbations in the correlation functions are 
damped and smoothened out by the evolution, although they do not reach exact thermal equilibrium.
In Fig.\ref{figure:5_1_2_20avst} we show the time evolution of $A(0)$ 
and $A(\Lambda/2)$. We remind that $\sqrt{TA(0)}$ should approach 
$\omega(0)$ in thermal equilibrium.
 
\section{Asymptotic behavior and thermalization}
\begin{figure}
\vspace{-5.5cm}
\centerline{\psfig{file=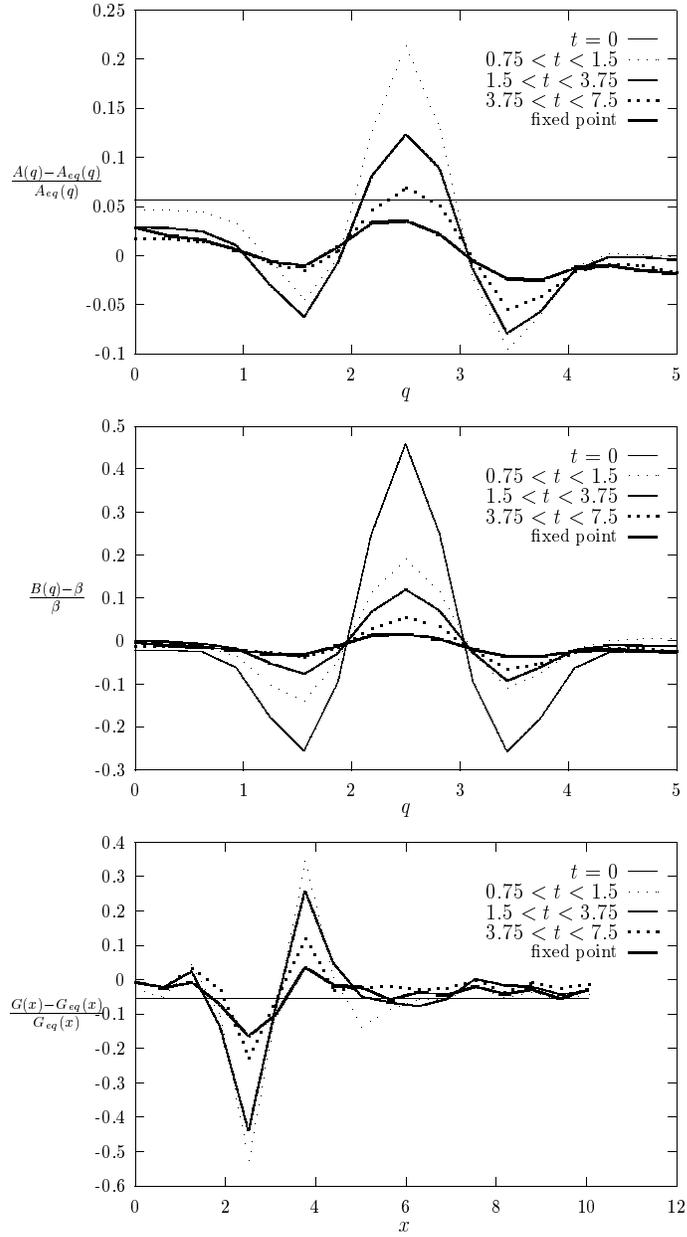,height=21cm}}
\caption{Time evolution of the two-point functions for a system with
  $\lambda=10$, $N=1$, $\beta=0.3$, with nonthermal initial conditions. The
  three panels show, from top to bottom,  the relative deviations of $A(q)$,
  $B(q)$ and $G(x)$, averaged over various time intervals, from their respective thermal values.}
\label{fig:ab_60_1}
\end{figure}

We have seen in the previous section that next-to-leading order terms in the
weighted loop or $1/N$ expansions
induce an energy exchange between different Fourier modes, which is a
prerequisite  for thermalization. Also "particle numbers" for individual
Fourier modes are no longer conserved separately.
Because of this energy exchange, a system with nonthermal initial conditions
is driven towards thermal equilibrium. 
At late times its correlation functions  oscillate around mean values 
that are ``more thermal'' than in the case of mere dephasing. 
In Fig.\ref{fig:ab_60_1} we show the evolution  of the time-averaged correlation functions $A(q)$ and $B(q)$ 
in a typical case. 
One clearly observes the initial approach towards the equilibrium values. 
For large $t$, however, stationary 
values are reached which deviate from the equilibrium correlations. These 
stationary values correspond to  
exact fixed points of the truncated evolution equations. 
We have computed the fixed points by methods similar to the computation 
of  the equilibrium. They are also displayed in 
Fig.\ref{fig:ab_60_1}. We notice that the effect of the deviation from 
thermal equilibrium  on the correlation  function in coordinate 
space is small, as can be seen from the plot of $G(x-y)$.

\begin{figure}
\vspace{-14cm}
\centerline{\psfig{file=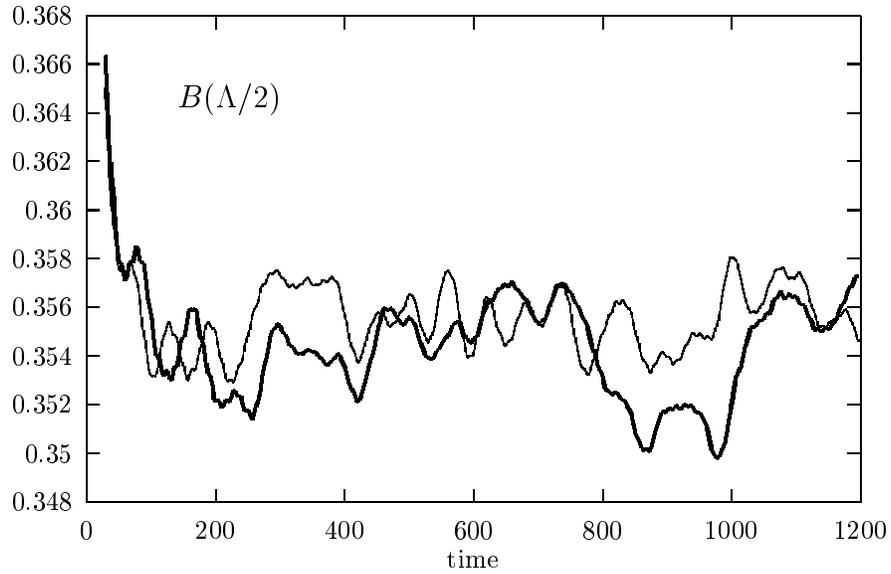}}
\vspace{-8cm}
\caption{Time evolution of $B(\Lambda/2)$ averaged over $\Delta t=30$. The parameters are
$\lambda=1$, $N=3$, $\beta=0.3$, $l=17.6$, and the initial perturbation
is described in the the caption of Table 1. The two curves correspond to the $1/N$ and weighted loop expansions (bold and plain, respectively). There is no sign or an asymptotic approach to the precise thermal value 
$B(\Lambda/2)=0.3$.}
\label{twotruncations}
\end{figure}

In Table 1 we collect the asymptotic stationary values  for several 
different choices of physical parameters and temperature, and otherwise
identical initial conditions\footnote{This holds up to discretization
  corrections, since we define identical initial conditions by identical
  continuous functions of $q$.}. They clearly differ from thermal equilibrium,
typically on the 10\% 
level for not too small values of $\lambda/N$.
As a general rule, the larger $\lambda/N$, the faster the system approaches
its  asymptotic limit, and the closer this limit is to the thermal values. In order to save on
computer time, we have therefore opted 
for rather small values of $N$ ($1<N<5$) and values of $\lambda$ between $1$
and  $60$. Of course, for small $N$ and/or large 
$\lambda/N$ the applicability of a $1/N$ or weigthed-loop expansion is
 questionable. There is no satisfactory way to 
assess the reliability of either truncation in the strong coupling  regime. 
However, we feel comforted  by the fact that the 
4-point functions never grow so large as to make the system unstable, and that
the time fluctuations of $L^2$
(which we remind is conserved exactly by the exact evolution but only up to
relative  corrections $\sim 1/N^2$ by the truncated equations)
 are always quite small (at the 1\% 
level for $\lambda=60$, $N=1$). The most direct test seems to be a
comparison between the results from the weighted-loop and the $1/N$ expansions.
For $N=1$, large $\lambda$ (Table 1b), the two truncations lead to very different large time stationary values, and it 
is therefore conceivable that the asymptotic departure from thermal 
equilibrium is  due to truncation errors. As $N$ 
gets larger and $N$ smaller, however, the two expansions agree much better 
(Fig.\ref{twotruncations}) and it is less  plausible
that higher order $1/N$ corrections could account for the asymptotic 
departures from equilibrium. We 
therefore believe that our numerical results support the existence of nonthermal attractive fixed points also for the 
exact ({\em i.e.} non-truncated) system. The implications of this claim are 
discussed in the next section.

As a final word of caution, we should consider the possibility that full thermalization {\em does} occur even for small
$\lambda/N$, but on longer time scales than those we have been able to probe. 
This issue can be settled only
with further, more computer-intensive investigations. What we have seen 
so far does not support this 
hypothesis. Even if that turned out to be the case, our method would still be
very useful for identifying the 
various time scales (dephasing - partial thermalization - complete thermalization) and for assessing the role of 
thermalization in practical problems, where  extremely large time scales
 are not always relevant.

\begin{figure}
\vspace{-10cm}
\centerline{\psfig{file=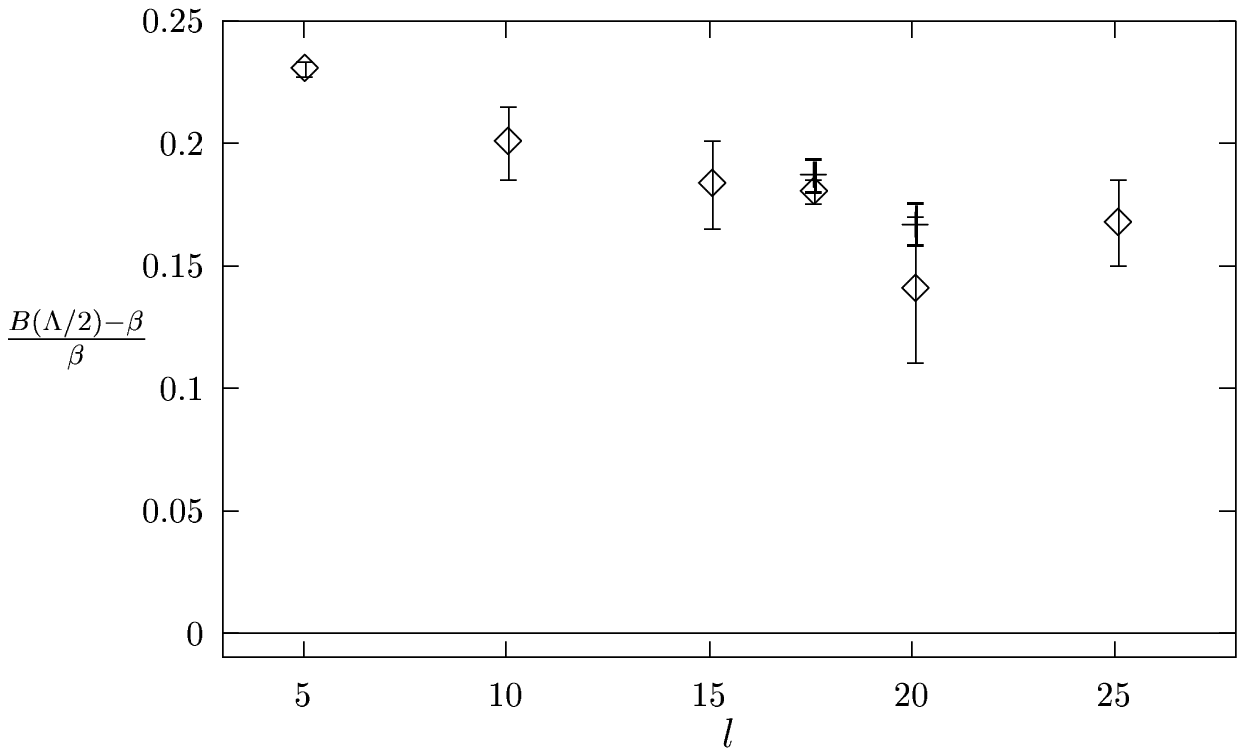}}
\vspace{-10cm}
\caption{Volume dependence. We show the asymptotic time-averaged value of $B(\Lambda/2)$  as a function of
 $l$, for the system of Table 1c. Diamonds and crosses correspond to the $1/N$ and weighted loop 
expansions, respectively.}
\label{fig:volume}
\end{figure}

We have also studied the volume dependence of the large time behavior for a 
particular choice of parameters (Table 1c, Fig.\ref{fig:volume}). Apparently,
the large volume limit still differs from thermal equilibrium. Relatively small 
volumes ($l\sim 20m^{-1}$) seem often to be  sufficient for an extrapolation to infinite volume.

Finally, we have investigated  a limited sample of initial conditions that are
 not symmetric under time reversal,
 and whose backward and forward evolutions therefore differ in their
 microscopic details. In all cases, the 
large time asymptotic averaged values of the correlation functions come out 
 the same in both time directions.
\vskip 1cm

\begin{tabular}{rccccc}
$\lambda$&$N$&$\beta$&$l$&$\frac{B(\Lambda/2)-\beta}{\beta}$&
$\frac{A(\Lambda/2)-A_{eq}(\Lambda/2)}{A_{eq}(\Lambda/2)}$\\\\
(*) 3&1&0.3&20.1&0.083$\pm$0.005&0.08\\
(*) 1&2&0.3&20.1&0.158$\pm0.005$&0.16\\
(*) 1&5&0.3&20.1&0.16$\pm$0.01&0.16\\\\

10&1&0.3&20.1&0.02$\pm 0.005$ &0.03\\
(*) 10&1&0.3&20.1&0.03$\pm$0.005 &0.03\\
60&3&0.3&20.1&0.07$\pm$0.01&0.07\\
60&1&0.3&20.1&-0.013$\pm$ 0.003&0.013 \\
(*) 60&1&0.3&20.1&0.008$\pm$ 0.003 &0.009\\\\

(*) 1&3&0.3&5.03&0.23$\pm 0.003$&0.23\\
(*) 1&3&0.3&10.05&0.20$\pm$ 0.015&0.20\\
(*) 1&3&0.3&15.08&0.183$\pm$ 0.015&0.24\\
(*) 1&3&0.3&17.6&0.18 $\pm$ 0.005&0.18\\
   1&3&0.3&17.6&0.187$\pm$ 0.006&0.185\\
(*) 1&3&0.3&20.1&0.11$\pm$ 0.03&0.12\\
1&3&0.3&20.1&0.167$\pm$ 0.008&0.17\\
(*) 1&3&0.3&25.1&0.167$\pm 0.017$&0.16\\

\end{tabular}
\vskip 0.5cm
{\small Table 1: Asymptotic displacement from thermal equilibrium  for 
different $\lambda$, $N$, $\beta$ and $l$.
For all configurations $\Lambda=5$.
The initial perturbation is a superposition of three gaussians with
$D_B=\beta/2$,  $q_B=\Lambda/2$, 
$\Delta_B=1.5\Lambda/16$; $D_B=-\beta/4$, $q_B=5\Lambda/16$, 
$\Delta_B=1.5\Lambda/16$; and $D_B=-\beta/4$, $q_B=11\Lambda/16$, 
$\Delta_B=1.5\Lambda/16$. 
Configurations marked by (*) are evolved according to the  $1/N$ 
expansion, the others according to the weighted loop expansion. For the first entries (a) the $1/N$
expansion seems reliable. The second group of entries (b) concerns large interactions with a rapid approach to asymptotic behavior. Finally, the last entries (c) are used for a study of the volume dependence, see also 
Fig.\ref{fig:volume}.}

\section{Discussion}

Our study of the evolution equations, applied to various initial nonthermal 
probability distributions, clearly establishes effectively irreversible behavior.
This is not put in by hand in the form of irreversible evolution equations. Our
equations are manifestly invariant under time 
reflection. Effective irreversibility is rather related to the existence of
 stationary solutions or fixed points towards which the 
flow is effectively attracted. It can be observed by evolving in  time
directions $t\rightarrow \pm \infty$. We see both the 
effects of dephasing, {\em i.e.} effective loss of phase information, and
scattering, {\em i.e.} energy exchange between different 
momentum modes. Our investigations are carried out for translation invariant
ensembles such that the energy exchange is not merely 
due to a classical background field evolving in time. The inclusion of
scattering  effects is a crucial step beyond the leading 
$1/N$-approximation used in the past \cite{1overN}. Genuinely, the system
 approaches asymptotically for large $t$ an oscillatory 
behavior of the correlation functions around a stationary solution. The time
 averaged values of the correlation functions are  
close to the corresponding stationary solutions. In a rough sense, the
stationary  solutions share many properties of thermal 
equilibrium. The corresponding fixed points are, nevertheless, not identical to the thermal fixed point.
The latter turns out to be a point  in a whole manifold of fixed points. For generic fixed points in this manifold the correlation functions differ from thermal equilibrium.

Perhaps the most interesting observation concerns the difference of the
asymptotic  stationary ensembles from the thermal ensemble. The system retains
memory of the initial conditions beyond the energy density 
or temperature. Isolated systems seem to differ in this respect from systems
coupled to a heat bath. 
Arguments why isolated systems do not thermalize exactly can be given on
different  levels. 
First there are exact obstructions from conserved correlation functions. An
example is the squared 
angular momentum density $<L^2>/l$. In thermal equilibrium this quantity can be computed as a function 
of temperature, $L^2_{eq}/l=N(N-1)T\int_qG_{eq}(q;T)$. Since $<L^2>$ is
conserved  by the exact flow  equations, any initial value of $<L^2>$
different  from the thermal one implies immediately that the correlation
functions  appearing in Eqs.(\ref{eq:energy})-(\ref{eq:momentum}) cannot all
take thermal  values for $t\rightarrow \infty$, not even in a time-averaged
sense. We emphasize that this
obstruction  is based on an exactly 
conserved quantity and therefore cannot be an artifact of insufficient
approximations. 

In principle, 
one could take care of the conservation of $<L^2>$ by an extension of the
thermodynamic description, adding a chemical potential for $L^2$. The problem
is,  however, that there exist infinitely many  conserved combinations of
correlation functions. Another
prominent example is the global ``specific heat'' $c_V=<(E-<E>)^2>/\;T^2l$ which 
corresponds again to an exactly conserved combination of correlation functions 
$<(E-<E>)^2>=\frac{Nl}{2}\int_q G^2_{\pi\pi}(q)+...$. Also, the additional
``chemical potentials''  would multiply nonlocal expressions like
$E^2=\int dxdy\;\epsilon(x)\epsilon(y)$.
Indeed, the  probability distribution $p\sim e^{-b(E/l)E}$ is stationary for an arbitrary function $b(\epsilon)$.
The Boltzmann distribution
 $b(\epsilon)=\beta$ is only a special case. If the function $b(\epsilon)$ is not  constant
 the  value of $c_V$ typically
 differs from the one in thermal equilibrium\footnote{The infinite volume behavior 
$(<E^2>-<E>^2)/<E^2>\sim l^{-1}$ holds for a wide class of $b(\epsilon)$ if its
deviation from a constant 
scales properly with $l$.}. These general considerations hold for an arbitrary 
number of space dimensions. 
They show that for isolated systems there cannot be a proof of strict thermalization\footnote{By ``strict
thermalization'' we mean an asymptotic 
approach of the probability distribution to the Boltzmann distribution, the distribution of the microcanonical ensemble.} using arguments of ergodicity. Strict thermalization for arbitrary initial conditions is in contradiction with the existence of conserved 
combinations of correlation 
functions\footnote{The general problem with ergodicity arguments is that only
 a finite neighborhood of a 
given point in phase space will be reached by an arbitrary trajectory after a 
finite lapse of time. This is not enough since even very close trajectories 
typically separate substantially at later time.}. 

On a second level one observes the existence of a large manifold of fixed
points  or stationary solutions 
for the truncated flow equations \cite{lb-cw2}. In our truncation they are
given  by $C=v=y=0$ and by 
solving the remaining equations $\partial_t C=\partial_t v=\partial_t
y=0$. The  latter equations determine the stationary values for $A$, $B$, $u$,
$w$, $s$, $z$ only incompletely. The present work clearly 
establishes numerically that these fixed points differ, in general, from the
one  corresponding to thermal 
equilibrium. They also prove stable with respect to small fluctuations. By 
investigating initial  conditions with the same $<L^2>$ as for thermal
equilibrium  we also establish explicitly that the fixed 
points are not not fully specified by $<E>$ and $<L^2>$. Numerically, we
actually  find a large manifold of
 different fixed points for given $<E>$ and $<L^2>$, as suggested by the
 counting  of equations and 
variables for the general stationary solutions. We also establish that the 
nonthermal fixed points play a 
role in the asymptotic dynamics. Nonthermal initial conditions typically
result in fluctuations around nonthermal stationary solutions at late time.

One may ask if the existence of these fixed points could not be an artifact of
 the  truncation. Three 
arguments indicate that this is not the case. First, some coordinates in the
 fixed point manifold are 
related to exactly conserved combinations of correlation functions like
 $<L^2>/l$.  Second, the counting of equations and variables 
indicates that the dimension of the fixed point manifold further increases
 once  1PI six-point functions or higher couplings are included. Third, we
 find  similar fixed points for different truncations in next-to-leading order in $1/N$. The ones approached by a given initial condition are close to each other for small $\lambda/N$ and stay substantially away from the thermal fixed points.

Our 
investigation of the volume dependence indicates that the fixed point manifold
does not shrink to the thermal fixed point in the infinite volume
limit. Furthermore, all numerical results suggest that the nontrivial fixed
points play indeed a dynamical role. From
all  this a picture for the  asymptotic late time  behavior of isolated
systems emerges where some features of "mesoscopic" dynamics survive even in 
the infinite volume limit.  The initial information is not lost completely as 
for a thermalizing  system. Part of the information 
survives and specifies the stationary solution around which the system
 oscillates asymptotically.

We observe that for large enough interactions the deviations from thermal
 equilibrium are small -
typically 
a few 
percent for the correlation functions and even less for quantities which
involve
 momentum averages. 
Part of this can be explained by exact relations which hold for {\em all}
stationary solutions. As an 
example, let us consider the condition for a static $<qp>$ correlation 
\ba
&&\frac{1}{2}\partial_t\int dx <q_a(x)p_a(x)>=\frac{1}{2}\int dx <p_a(x)p_a(x)>\\
&&-\frac{1}{2}\int dx\left\{ <\partial_i q_a(x)\partial_i q_a(x)>+m^2<q_a(x)q_a(x)>
\right.\\
&&\left.+\frac{\lambda}{2N}<(q_a(x)q_a(x))^2>\right\}=0\nonumber
\ea
which relates the kinetic and potential energy (note that they are not equal
for 
interacting systems):
\be
<E_{kin}>=<E_{pot}>+\frac{\lambda}{8N}\int dx <(q_a(x)q_a(x))^2>. \label{eq:ekep}
\ee
This is, of course, the thermal relation, but it extends to all other
stationary solutions as well. We conjecture that large interacting systems
 generically   show an effective irreversible evolution  towards 
asymptotic oscillations around one of the stationary solutions. Then relations
of  the type (\ref{eq:ekep})
 hold asymptotically irrespectively of the initial conditions. This explains
 the robustness of a large set of asymptotic
 time averages of correlation functions - an important part of the initial 
information is indeed lost.
 Conversely, a judgement of precise asymptotic thermalization should not be
 based  on generic relations  like Eq.(\ref{eq:ekep}), but rather on
 correlation  functions which can differ for two inequivalent fixed points.

The lack of exact thermalization of large interacting systems has consequences
 for
 ``systems in a heat bath'' 
as well. Indeed, we may consider a subsystem, say $q_a(x)$, $p_a(x)$ for
 $|x|\leq 
l_0/2\ll l/2$ and view it
 as evolving in the ``heat bath'' consisting of the degrees of freedom with 
$l_0/2<|x|\leq l/2$. Can we expect 
that the subsystem effectively thermalizes even though the large isolated
 system 
(subsystem and bath) does not 
exactly thermalize? This question can be addressed by an investigation of 
correlation functions for the 
subsystem, say  $<q_a(x)q_a(y)>_c$ with $|x|, |y|\leq l_0/2$, or a convenient 
smoothened version ($k_0=\pi/l_0$)
\be
G_{k_0}(x,y)=\frac{1}{N}<q_a^{k_0}(x)q_a^{k_0}(y)>_c=G(x-y)e^{-\frac{1}{2}k_0^2(x^2+y^2)}.
\ee
\be
q_a^{k_0}(x)=q_a(x)e^{-\frac{1}{2}k_0^2x^2}
\ee
For a translationally invariant ensemble one has 
\ba
G_{k_0}(x,y)&=&\int_{q,q'} e^{i(qx-q'y)}G_{k_0}(q,q')\\
G_{k_0}(q,q')&=&\left(\frac{2\pi}{k_0^2}\right)^D\exp\left(-\frac{(q-q')^2}{4k_0^2}\right)\cdot\nonumber\\
&&\int_pG(p)\exp\left(-\frac{(p-\frac{1}{2}(q+q'))^2}{k_0^2}\right)\nonumber
\ea
In particular, the Fourier transform of $G_{k_0}(x,0)$ reads
\be
\tilde G_{k_0}(q)=\int_{q'}G_{k_0}(q,q')=(2\pi)^{D/2}k_0^{-D}\int_p G(p)e^{-\frac{(q-p)^2}{2k_0^2}}.\label{eq:AA}
\ee
Only for $k_0\rightarrow 0$ does this reduce to $G(q)$. For nonzero $k_0$
(corresponding to a subsystem), however, 
$\tilde G_{k_0}(q)$ involves a momentum averaging with width given by $k_0$. 
Because of  dephasing, the momentum 
averaged two-point function $\tilde G_{k_0}(q)$ approaches a stationary value
 much more efficiently than $G(q)$.
 For large $l_0$, and nevertheless $l_0\ll l$, it is conceivable that 
$\tilde G_{k_0}(q)$ actually reaches 
asymptotically a stationary value whereas $G(q)$ fluctuates around a
 stationary 
value.
Nevertheless, the time averaged   values of $\tilde G_{k_0}(q)$ and $G(q)$ are 
still related by Eq.(\ref{eq:AA}).
A nonthermal asymptotic behavior of the time averaged $G(q)$ will manifest
 itself 
also in the asymptotic form of
$\tilde G_{k_0}(q)$ if it extends over a momentum range with width larger than 
$k_0$. Only variations of $G(q)$ 
in small momentum ranges will be washed out. From our present investigation we
 see
 no indication that asymptotically
$\tilde G_{k_0}(q)$ reaches precisely its thermal value. We conclude that
 mesoscopic
 dynamics may also be of 
relevance for subsystems which are in thermal contact with a ``heat
 bath''. The 
crucial point here is that the 
heat bath itself is not precisely thermalizing.

In summary, our investigation indicates that isolated systems roughly
thermalize 
for large time, while some quantitative deviations 
from thermal equilibrium remain. The ``loss of memory of the initial
conditions'',
 usually assumed in the picture of thermalization,
 turns out not to be complete. This holds for interacting systems and in the
 large
 volume limit. Our results question Boltzmann's 
thermalization conjecture for isolated systems. They suggest that even large 
interacting isolated systems do not thermalize in a 
strict sense.

Our treatment is based on an exact evolution equation for the time dependence
of 
equal time correlation functions. Nevertheless, 
the solution of these equations involves approximations in the form of a 
truncation of the time dependent effective action. Since our findings touch the basics of 
thermodynamics, they should be questioned by an 
independent method. One possibility seems the 
numerical solution of the microscopic equation (\ref{eq:micro}) for a large
sample
 of different initial conditions. Taking an 
ensemble average over the initial conditions gives directly the equal time 
correlation functions which can be compared with the 
present work. Such a computation could establish definitely if the findings of 
this work are substantially affected by the 
truncation or not.

{\bf Acknowledgement:} We thank L. Bettencourt for many helpful discussions.

\section{Appendix A: Flow Equations}
In this appendix we present the evolution equation for the 1PI 4-point
functions which are not specified in the main text.
\ba
   \dot v(q_1,q_2,q_3)&=&\left[
        2\omega^2(\2)(w(\1,\2,\3)-s(\1,\2,\3)-s(\1,\2,\4))
        \right.\nonumber\\
      &&+2\omega^2(\3)s(\1,\3,\2)-4u(\1,\2,\3)+4\lambda B(\1)
        -\frac{\gamma(\1)}{N}v(\1,\2,\3)
         \nonumber\\
        &&-4\lambda (B(\1)(S_1(\1+\2,\3)+S_2(\1+\3,\2))+
        C(\4)S_3(\1+\2,\1)\nonumber\\
        &&+2C(\2)S_5(\2+\3,\1)+C(\4)S_6(\1+\3,\1))\nonumber\\
        &&\left.-\lambda\hbar^2B(\1)(C(\2)C(\3)+
        C(\3)C(\4)+C(\2)C(\4))\right]_{SYM}\nonumber\\  
  \dot w(q_1,q_2,q_3)&=&\left[\omega^2(\3)(y(\1,\2,\3)+y(\1,\3,\2)+y(\2,\3,\1))
     -v(\1,\2,\3) \right.\nonumber\\
     &&-v(\2,\4,\1)-v(\2,\3,\1)-\frac{\gamma(\1)+\gamma(\2)}{N}w(\1,\2,\3)
     \nonumber\\
     &&-\lambda(C(\3)S_4(\1,\2)+8B(\2)S_5(\2+\3,\1))\nonumber\\
        &&-4\lambda(B(\1)S_3(-\1-\3,\2)+
     B(\1)S_6(\2+\3,\2) \nonumber\\
     &&\left.+C(\3)S_7(\1,\2))-3\lambda\hbar^2B(\1)B(\2)C(\3)\right]_{SYM}\nonumber\\
   \dot s(q_1,q_2,q_3)&=&\left[ 2\omega^2(\3)y(\1,\3,\2)-2v(\2,\4,\1)
   -\frac{\gamma(\1)+\gamma(\2)}{N}s(\1,\2,\3)\right.\nonumber\\
   &&-4\lambda(B(\1)S_3(-\1-\3,\2)+B(\1)S_6(\2+\3,\2)+C(\3)S_7(\1,\2))\nonumber\\
    &&\left.-2\lambda\hbar^2B(\1)B(\2)C(\3)
\right]_{SYM}\nonumber\\
   \dot y(q_1,q_2,q_3)&=&\left[
   4\omega^2(\4)z(\1,\2,\3)-2w(\1,\2,\3)-s(\2,\3,\1)
   -s(\1,\3,\2)\right.\nonumber\\
 &&+s(\1,\2,\3)+s(\1,\2,\4)-\frac{\gamma(\1)+\gamma(\2)+\gamma(\3)}{N}
 y(\1,\2,\3)\nonumber\\
 &&\left.-4\lambda(B(\3)\frac{S_4(\1,\2)}{4}+B(\1)S_7(\2,\3))-\lambda\hbar^2B(\1)B(\2)B(\3)\right]_{SYM}
\nonumber\\
   \dot z(q_1,q_2,q_3)&=&\left[-y(\1,\2,\3)-4\frac{\gamma(\1)}{N}z(\1,\2,\3)
   \right]_{SYM}.
\ea

They involve the following momentum integrals:
\ba
S_3(\1,\2)&\equiv&\frac{1}{2N}\int_q G(q)G(q-\1)
\left[\frac{1}{2}v(\2,-q,q-\1)+\frac{1}{2}v(\2,q-\1,-q)
   \right.\nonumber \\
&&+\frac{N+2}{2}v(\2,\1-\2,-q)+ \frac{1}{2}c(q)c(q-\1)((N+2)y(q-\1,-q,\2)\nonumber\\
&&+y(\2,q-\1,-q)+y(\2,-q,q-\1))-c(q)s(-q,\2,q-\1)\nonumber\\
&&\left.-c(q-\1)(Ns(q-\1,\2,-q)+2w(q-\1,\2,-q))\right]\nonumber\\
S_4(\1,\2)&\equiv&\frac{1}{N}\int_q G(q)G(q-\1-\2)\cdot \nonumber\\
&&\left[(N+2)(w(\1,\2,q-\1-\2)-\frac{1}{2}s(\1,\2,q-\1-\2)-\frac{1}{2}s(\1,\2,-q))\right.\nonumber\\
&&+s(\1,\2,-q)+2c(q-\1-\2)c(q)(Nz(q-\1-\2,-q,\1)\nonumber\\
&& +2z(\2,q-\1-\2,\1)+2z(\2,\1,q-\1-\2))\nonumber\\
&& \left.-c(q)(Ny(\1,\2,-q)+2y(-q,\1,\2)+2y(\1,\2,-q))\right]\nonumber\\
S_5(\1,\2)&=&\frac{1}{8N}\int_q G(q)G(-q-\1)\cdot \nonumber\\
&&\left[2v(\2,-q,q+\1)+2c(q)c(q+\1)y(\2,-q,q+\1)\right.\nonumber\\
&&-4c(q)(w(-q,\2,q+\1)-\frac{1}{2}s(-q,\2,q+\1)
-\frac{1}{2}s(-q,\2,-\1-\2))\nonumber\\
&&\left.-2c(-q-\1)s(\2,q+\1,-q)
\right]\nonumber\\
S_6(\1,\2)&=&\frac{1}{2N}\int_q G(q)G(-q-\1)\cdot \nonumber\\
&&\left[v(\2,-q-\1,q)-c(q)s(\2,-q,-q-\1)-2c(-q-\1)(w(-q-\1,\2,q)
\right.\nonumber\\
&&-\frac{1}{2}s(-q-\1,\2,q)-\frac{1}{2}s(-q-\1,\2,\1-\2))\nonumber\\
&&\left.
+c(q)c(-q-\1)y(\2,-\1-q,q)\right]\nonumber\\
S_7(\1,\2)&=&\frac{1}{2N}\int_q G(q)G(q-\1-\2)\cdot \nonumber\\
&&\left[s(\2,\1,-q)-c(q)y(-q,\2,\1)-c(q-\1-\2)y(q-\1-\2,\1,\2)\right.\nonumber\\
&&\left.+4c(q)c(q-\1-\2)z(\1,q-\1-\2,\2)\right].
\ea

\end{document}